\documentclass[aps,prd,twocolumn,groupedaddress]{revtex4-1}
\usepackage{graphicx}
\include{amssym}

\begin{document}
\title{Proper-time method for unequal masses}
\author{A. A. Osipov\footnote{Email address: osipov@nu.jinr.ru}}
\affiliation{Joint Institute for Nuclear Research, Bogoliubov Laboratory of Theoretical Physics, 141980 Dubna, Russia}

\begin{abstract}
The result of removing of heavy non-equal mass particles from the theory can be described, at low energy, by the effective action, which is a series in inverse-square powers of the mass. We propose a new efficient tool to calculate the leading terms of this series based on the Schwinger proper-time method. Unequal masses give rise to a large number of effective vertices describing the explicit flavour symmetry breaking effects with well-defined coupling constants. Our method is pertinent to the theory with explicit and spontaneous chiral symmetry breaking, chiral gauge theory, standard and beyond standard model effective field theory, the theory of critical phenomena, cosmology, etc.
\end{abstract}

\maketitle

\vspace{0.5cm}

{\it Keywords: proper-time method, heat kernel, effective field theory, effective action, flavor symmetry breaking, unequal masses.}

\section{Introduction}
The proper-time method \cite{Fock:37a,Fock:37b,Nambu:50,Schwinger:51} is an efficient tool to study quantum corrections in many areas of theoretical physics: in quantum gravity \cite{DeWitt:65,DeWitt:75,Barvinsky:87,Barvinsky:90}, in QCD (effective meson Lagrangians \cite{Wadia:85,Ebert:86}), in chiral gauge theories \cite{Ball:89}, in cosmology \cite{Callan:94}, in QED (Casimir energies and forces \cite{Bordag:01}), etc. That is why it remains in the focus of researchers working in quantum physics for decades \cite{Vassilevich:03}. 

Here I propose a new proper-time based algorithm to derive the effective action in the theory with heavy virtual fermions (or bosons) of unequal masses belonging to some representation of the symmetry group $G$. The result is the inverse mass series for the one-loop effective action, where I explicitly calculate the two leading contributions. This powerful tool opens a promising avenue for studying explicit flavor symmetry breaking effects in many effective field theories. It provides the easiest way to address a problem that otherwise becomes rather cumbersome. 

To generalize the standard large mass expansion of the heat kernel to the case of unequal masses, I suggest using the formula
\begin{equation}
\label{alg}
e^{-t(M^2+A)}=e^{-tM^2}\left[1+\sum_{n=1}^\infty (-1)^n f_n(t,A) \right],
\end{equation}
where $M=\mbox{diag} (m_1, m_2, \ldots ,m_f)$ is the diagonal mass matrix; $t$ is the proper-time parameter; the expression in the square brackets is the time-ordered exponential $\mbox{OE}[-A](t)$ of $A(s)= e^{sM^2} A e^{-sM^2}$, and $A$ is a positive definite self-adjoint elliptic operator in some background (its explicit form will be clarified later), accordingly 
\begin{equation}
f_n(t,A)=\int\limits_0^t\!\! ds_1\!\!\int\limits_0^{s_1}\!\! ds_2 \ldots \!\!\!\!\int\limits_0^{s_{n-1}}\!\!\!\! ds_n A(s_1) A(s_2) \ldots A(s_n).
\end{equation} 
If masses are equal, this formula yields the well-known large mass expansion with standard Seeley-DeWitt coefficients $a_n(x,y)$ \cite{Ball:89}. These coefficients are polynomials in background fields and describe, in the coincidence limit $y\to x$, local vertices of the induced effective Lagrangian. In fact, formula (\ref{alg}) is an extension of the Schwinger's method used to isolate the infinities contained in the real part of the one-loop action \cite{Schwinger:51,DeWitt:75} to the non-commutative algebra. 

There is a simple heuristic argument that explains why this formula is also relevant for describing the generalized $1/M$ series. Indeed, the $1/M$ expansion is known to be valid when all background fields and their derivatives are small compared to the mass of quantum fields. Therefore, factoring $e^{-tM^2}$ one separates the leading contribution. The remaining part of the heat kernel may be unambiguously evaluated by expanding it in a power series in $t$ about $t=0$. As a consequence, the Seeley-DeWitt coefficients $a_n$ receive corrections $a_n\to b_n=a_n+\Delta a_n$, where $\Delta a_n$ vanish in the limit of equal masses.  

Currently, there are two methods for deriving quantum corrections induced by virtual states of unequal masses. In \cite{Min:82,Min:89a,Min:89b}, the heat kernel is evaluated on the bases of the modified DeWitt WKB form. 
This yields a different asymptotic series for the right-hand side of Eq.(\ref{alg}), and, consequently, the different expressions for $\Delta a_n$. The approach proposed in \cite{Osipov:01a,Osipov:01b,Osipov:01c} starts from the formula (\ref{alg}), but afterwards an additional resummation of the asymptotic series is applied. This essentially simplifies the calculations, but changes the structure of the $1/M$ series. As a result, one looses correspondence between a mass-dependent factor at the effective vertex and a flavor content of the one-loop Feynman diagram which generates the vertex. Here I abandon this procedure.

The proper-time method reduces the task of the large mass expansion to a simple algebraic problem which requires much less work than one needs for the corresponding Feynman diagrams calculation in momentum space. In the following, we consider a quite nontrivial case of the chiral $U(3)\times U(3)$ symmetry broken by the diagonal mass matrix $M=\mbox{diag} (m_1, m_2, m_3)$ to demonstrate the power of the method. To find the two leading contributions $b_1$ and $b_2$ in the $1/M$ expansion of the effective action one requires to consider only four terms of the series (\ref{alg}) that results in more than a hundred effective vertices. 

The effects of flavor symmetry breaking are currently important in many physical applications: in studies of physics beyond standard model to construct the low energy effective action by integrating out the heavy degrees of freedom \cite{Passarino:19,Bizot:18}; in two Higgs doublet models \cite{Branco:12} to address the problem of almost degenerate Higgs states at $125 \mbox{GeV}$ \cite{Bian:18,Haber:19}; in the low energy QCD to study the $SU(3)$ and isospin symmetry breaking \cite{Taron:97}. These effects are very important to address the QCD phase diagram \cite{Osipov:13}, to study a formation of the strange-quark matter \cite{Witten:84,Osipov:15}, to study nuclear matter in extreme conditions that arose in nature at the early stages of the evolution of the Universe and in the depths of neutron stars \cite{Chernodub:11,Gatto:11}.

\section{Proper-time expansion}
We shall be mainly concerned here with the asymptotic behavior of the following object in Euclidean four-dimensional space
\begin{equation}
\label{logdet}  
  W_E =-\ln |\det D_E| 
         =\!\!\int\limits^\infty_0\!\frac{dt}{2t}\,\rho_{t,\Lambda}\,\mbox{Tr}\left(e^{-t D_E^\dagger D_E^{}}\right),
\end{equation}
It represents the real part of the one-quark-loop contribution to the meson effective action in form of the proper-time integral. The integral diverges at the lower limit, therefore, a regulator $\rho_{t,\Lambda}$ is introduced, where $\Lambda$ is a cutoff. The Dirac operator in Euclidean space, $D_E$, has the form  
\begin{equation}
D_E^{}=i\gamma_\alpha d_\alpha-M+ s+i\gamma_{5} p ,
\end{equation}
where $d_\alpha=\partial_\alpha +i\Gamma_\alpha$, $\Gamma_\alpha =v_\alpha +\gamma_{5}a_\alpha$, $\alpha =1,2,3,4$. The external scalar $s$, pseudoscalar $p$, vector $v_\alpha$ and axial-vector $a_\alpha$ fields are embedded in the flavor space through the set of matrices $\lambda_a = (\lambda_0, \lambda_i)$, where $\lambda_0 =\sqrt{2/3}$ and $\lambda_i$ are the eight $SU(3)$ Gell-Mann matrices; for instance,  $s=s_a\lambda_a$, etc. for all fields. The quark masses are given by the diagonal matrix $M=\mbox{diag} (m_1, m_2, m_3)$ in the flavor space. The symbol "Tr" denotes the trace over Dirac $(D)$ $\gamma$-matrices, colour $(c)$ $SU(3)$ matrices, and flavor $(f)$ matrices, as well as integration over coordinates of the Euclidean space: $\mbox{Tr}\equiv \mbox{tr}_I \int\! d^4x $, where $I=(D,c,f)$. The trace in the color space is trivial: it leads to the overall factor $N_c=3$. The dependence on external fields in $D_E$ after switching to the Hermitian operator
\begin{equation}
D_E^\dagger D_E^{}=M^2 -d_\alpha^2+Y
\end{equation}
is collected in $Y$ and the covariant derivative $d_\alpha$. In the following we do not need the explicit form of $Y$.

To advance in the evaluation of expression (\ref{logdet}), we use the Schwinger technique of a fictitious Hilbert space \cite{Schwinger:51}. It allows the effective action to be represented as an integral over the 4-momenta $k_\alpha$
\begin{equation}
\label{logdet2}  
     W_E=\!\! \int\!\! d^4x\!\!\int\!\!\frac{d^4k}{(2\pi )^4}\, e^{-k^2}\!
            \!\! \int\limits^\infty_0\!\!\frac{dt}{2t^3}\,\rho_{t,\Lambda}\,
          \mbox{tr}_I \left[e^{-t(M^2+A)}\right]\! ,
\end{equation}
where $A = -d_\alpha^2 -2ik\partial / \sqrt{t} +Y $. Since $A$ contains open derivatives with respect to coordinates, it is still necessary to clarify the meaning of $\mbox{tr}_I$. The space $x\in {\bf R}^4$ has no boundaries, therefore we can integrate by parts. However, this operation will be unambiguous only if $\mbox{tr}_I$ does not change when its $x$-dependent elements are cyclically rearranged. This property can be ensured by the explicit symmetrization, i.e., $\mbox{tr}_I$ in such cases is understood as
\begin{equation}
\label{str}
\mbox{str}_I\, (A_1A_2\ldots A_n)=\frac{1}{n}\sum_{cycl. perm.}\!\!\!\!\mbox{tr}_I\, (A_{1}A_{2}\ldots A_{n})\,.
\end{equation}

To distinguish the leading contributions in the expansion in $1/M^2$, it suffices to take into account terms of at most $t^4$ order in (\ref{alg}), which under the flavor trace have the form
\begin{eqnarray} 
\label{sum2}
 &&\mbox{tr}_f\left[e^{-t(M^2+A)}\right]=\sum_{i=1}^3 c_i(t) -t\sum_{i=1}^3  c_i(t)\, \mbox{tr}_f\, A_i \nonumber\\ 
 &+&\frac{t^2}{2!}\sum_{i,j} c_{ij}(t) \, \mbox{tr}_f\, (A_i A_j) -\frac{t^3}{3!}\sum_{i,j,k} c_{ijk} (t)\, \mbox{tr}_f\, (A_iA_jA_k) \nonumber\\
& +&\frac{t^4}{4!}\sum_{i,j,k,l} c_{ijkl} (t)\, \mbox{tr}_f\, (A_iA_jA_kA_l)+{\cal O}(t^5)\,. 
\end{eqnarray}
Here $A_i = E_i A$, where $3 \times 3$ matrices $(E_i)_{nm} = \delta_{in} \delta_{im}$; the coefficients $c_{i_1i_2 \ldots i_n}(t)$ are completely symmetric with respect to any permutation of indices and are well known \cite{Osipov:01a,Osipov:01b,Osipov:01c}. They depend on quark masses and proper-time. In the case of equal masses, all coefficients coincide, that is, $c_{i} (t) = c_{ii} (t) = c_{iii} (t) = c_{iiii} (t) = e^{ -t m_i^2}$. It also follows from the formula (\ref{sum2}) that the number of indices in the coefficient $c_{i_1i_2\ldots i_n}$ is equal to the number of operators $A_i$, representing external fields. Hence, there is a relation between the number of indices in $c_{i_1i_2\ldots i_n}$ and the contribution of the one-loop diagram with $n$ external legs. The formula (\ref{integrals}) will clarify this correspondence.

Substituting (\ref{sum2}) into (\ref{logdet2}) and integrating over momenta $k_\alpha$ and proper-time $t$, we arrive at the expression
\begin{equation}
\label{WE}
W_E=  \frac{N_c}{32\pi^2}\!\!\int\!\! d^4x \sum_{n=0}^{\infty}\,\mbox{tr}_{Df}\,b_n(x,x),
\end{equation}
where coefficients $b_n(x, x)$ depend on the external fields and quark masses, i.e., they contain information about both the effective meson vertices and corresponding coupling constants. If all masses are equal, the dependence on $m=m_i$ is factorized in form of the integral \cite{Osipov:01b}
\begin{equation}
\label{Jn}
J_n (m_i^2)=\int\limits^\infty_0  \frac{dt}{t^{2-n}}\,\rho_{t,\Lambda}\, c_{i} (t)
\end{equation}
and the field-dependent part takes a standard Seeley-DeWitt form $a_n(x,x)$. For large masses $m$, the coefficients $b_n(x, x)$ exhibit the same asymptotic behavior as $J_{n-1} (m^2)$, i.e., $b_n\sim m^{- 2 (n-2)}$. 

\section{Leading coefficients}
Consider the leading terms $b_1$ and $b_2$. The case $n = 0$ is of no interest because $b_0$ contains no fields and can be omitted from the effective action. The coefficients with $n\geq 3$ tend to zero in the limit of infinite masses, therefore they are small in comparison with $b_1$ and $b_2$. 

For convenience of writing the result of our calculations, along with the usual matrix multiplication, we will use the non-standard Hadamard product \cite{Styan:73}, which is the matrix of elementwise products 
\begin{equation}
(A\circ B)_{ij} =A_{ij} B_{ij}.
\end{equation}
The Hadamard product is commutative unlike regular matrix multiplication, but the distributive and associative properties are retained. It has previously been proven to be an useful tool when non-degenerate mass matrices and non trivial flavor symmetry contractions were involved \cite{Morais:17}.

Now, the heat coefficient $b_1(x,x)$ can be written as
\begin{equation}
b_1=  - J_0\circ Y, 
\end{equation}
where $(J_0)_{ij}=\delta_{ij} J_0(m_i^2)$ is a diagonal matrix with elements given by (\ref{Jn}) (for $n=0$). This matrix contains contributions of the Feynman one-loop diagram, known as a "tadpole". The regularization should be chosen in accord with the problem studied, for the NJL model this can be $\rho_{t,\Lambda}=1-(1+t\Lambda^2)e^{-t\Lambda^2}$.  

Consider the second coefficient $b_2(x,x)$. After some tiring calculations we can represent the result in the form 
\begin{equation}
\label{b2}
b_2=\frac{1}{2}\, Y\left(J\circ Y\right)-\frac{1}{12} \Gamma^{\alpha\beta} \left(J\circ \Gamma^{\alpha\beta} \right)+\Delta b_2,
\end{equation}
where the antisymmetric tensor $\Gamma_{\alpha\beta} = F_{\alpha\beta} + i[\Gamma_\alpha, \Gamma_\beta ] $, $F_{\alpha\beta} = \partial_\alpha \Gamma_\beta - \partial_\beta \Gamma_\alpha$, and $J_{ij} = J(m_i, m_j)$ is a symmetric $ 3\times 3 $ matrix, whose elements are logarithmically divergent parts (at $\Lambda \to \infty $) of Feynman self-energy diagrams with masses of virtual particles $m_i$ and $m_j$. Since we also need expressions for the similar contributions coming from triangular $J_{ijk} $ and box $J_{iijk}$ diagrams, we collect them together in one formula 
\begin{equation}
\label{integrals}
\left( \begin{array}{c}
J_{ij} \\  J_{ijk} \\  J_{iijk} \\
\end{array}\right)=\!\!\int\limits^\infty_0  \frac{dt}{t}\,\rho_{t,\Lambda}\, \left(
\begin{array}{c} c_{ij} (t) \\ c_{ijk} (t) \\ c_{iijk} (t) \\ \end{array}\right).
\end{equation}
In the theory under consideration, there are only three flavors of quarks, therefore in $J_ {iijk}$ at least two indices coincide. From the known properties of $c_{i_1i_2\ldots i_n}(t)$ one can deduce that $J_{ii}=J_{iii}=J_{iiii}$. The integrals take this form when the quark masses are equal. Their coincidence is an important property which is used to demonstrate that $\Delta b_2$ vanishes in the equal-mass limit.

In the limit of equal masses, the first two terms in (\ref{b2}) yield the well-known result \cite{Ball:89}, and the third one vanishes, that is, it contains only contributions associated with an explicit violation of chiral symmetry. Without this term, the description of exact breaking of flavor symmetry is incomplete. To write an expression for $\Delta b_2$, let us consider the different contributions in
\begin{equation}
\label{db2}
\Delta b_2=\Omega_Y +\sum_ {n=0}^{2} \Omega^{(n)},
\end{equation}
where $\Omega_Y$ is the sum of all terms linear in $Y$. $\Omega^{(n)}$ is the sum of terms with $n$ derivatives which consists of only spin-one fields. 

For $\Omega_Y$ we have 
\begin{equation}
\label{om1}
\Omega_Y= \mathcal AY+\frac{i}{3}(R\circ \Gamma^\alpha )\! \stackrel{\leftrightarrow}{\partial^\alpha}\! Y,
\end{equation}
where $R$ and $\mathcal A$ are $3\times 3$ matrices with elements  
\begin{eqnarray}
 R_{ij} &=& -\frac{1}{2} \left(  J_{iij}-J_{jji}  \right),    \\
 \mathcal A_{ii} &=& \sum_{j\neq i} \left(J_{ii}-J_{ij}+R_{ij}\right)\Gamma^\alpha_{ij} \Gamma^\alpha_{ji}\,, \nonumber \\
\mathcal A_{ij} &=& \left[(J_{ij}-J_{ijk})\Gamma^\alpha_{ik} \Gamma^\alpha_{kj} \right. \nonumber\\
&-&\left. R_{ij}\left(\Gamma^\alpha_{ij} \Gamma^\alpha_{jj}-\Gamma^\alpha_{ii} \Gamma^\alpha_{ij}\right)\right]_{\mid i\neq j\neq k}\,,
\end{eqnarray}
and the left-right derivative is defined by the difference $Y\! \stackrel{\leftrightarrow} {\partial^\alpha} \! \Gamma^\alpha = Y \partial^\alpha \Gamma^\alpha - (\partial^\alpha Y) \Gamma^\alpha$. To avoid misunderstandings, we emphasize that repeated flavor indices do not imply summation over them. Here and below, summation is carried out only if the summation symbol is explicitly written.

Let us consider the terms with two derivatives
\begin{equation}
\label{om2}
\Omega^{(2)}= \frac{1}{24} \left[F^{\alpha\beta}\left( T\circ F^{\alpha\beta}\right)
+6(\partial\Gamma )\left( T \circ \partial\Gamma\right)\right].
\end{equation}
The first term makes an additional contribution to the kinetic part of the effective Lagrangian of spin-1 fields described by the second term in ($\ref{b2}$). The symmetric matrix $T$ is defined by $T_{ij} = J_{ij} -J_{iijj} $. The shorthand $\partial\Gamma\equiv\partial^\alpha\Gamma^\alpha$  implies summation over omitted indices $\alpha$. In applications, one can omit the second term in (\ref{om2}) to ensure that the energy of the massive spin-1 field is positive definite. 

The terms with one derivative can be collected in the expression
\begin{eqnarray}
\label{om1}
\Omega^{(1)} 
&=& \frac{i}{3}C^{\alpha\beta} F^{\alpha\beta}+\frac{i}{6}\,\delta_{\alpha\beta\gamma\sigma}(K\circ \partial^\alpha\Gamma^\beta )E^{\gamma\sigma} \nonumber \\  
&+&\frac{3i}{4}\left( T \circ \partial^\alpha\Gamma^\beta\right)L^{\beta\alpha}_-,  
\end{eqnarray}
\linebreak
which represents the effective three-particle vertices describing the local interaction of vector and axial-vector fields. Symbol $\delta_{\alpha\beta\gamma\sigma} = \delta_{\alpha\beta} \delta_{\gamma\sigma} + \delta_ {\alpha\gamma} \delta_{\beta\sigma} + \delta_{\alpha\sigma} \delta_{\beta\gamma}$ is a totally symmetric tensor composed of the product of two Kronecker symbols. The diagonal and off-diagonal elements of the matrix $C^{\alpha\beta}$ are, respectively, of the form
\begin{eqnarray}
C^{\alpha\beta}_{ii}&=&\sum_{j\neq i} \left(J_{ii}-J_{ij}+\frac{1}{4}\,R_{ij}+\frac{3}{4}\, T_{ij} \right)\Gamma^{\alpha}_{ij}\Gamma^{\beta}_{ji}\,, \nonumber \\
C^{\alpha\beta}_{ij}&=&(J_{ij} -J_{123})\Gamma^{\alpha}_{ik}\Gamma^{\beta}_{kj}|_{i \neq j\neq k}
+\frac{1}{4}\, R_{ij} \left(L^{\alpha\beta}_+\right)_{ij} \nonumber \\
&+&\frac{3}{4}\, T_{ij} \left(L^{\alpha\beta}_-\right)_{ij},
\end{eqnarray}
where we use the notation
\begin{equation}
\left(L^{\alpha\beta}_\pm\right)_{ij}=\left(\Gamma^\alpha_{ii}\pm \Gamma^\alpha_{jj}\right)\Gamma^\beta_{ij}.
\end{equation}
Note that elements of the matrix $\Gamma^\alpha_{ij}$ commute with each other. The antisymmetric matrix $K$ is defined by the expression
\begin{equation}
K_{ij}=(J_{jjik}-J_{iijk})_{\mid i\neq j\neq k} \,.
\end{equation}
The tensor $E^{\gamma\sigma}$ has no diagonal elements too, although it is not antisymmetric
\begin{equation}
E^{\gamma\sigma}_{ii}=0, \quad
E^{\gamma\sigma}_{ij}=\Gamma^{\gamma}_{ik}\Gamma^{\sigma}_{kj}, \,\, (i\neq j\neq k). 
\end{equation}

The last thing left to consider is the term without derivatives $\Omega^{(0)}$, which is a sum of the terms $\propto\Gamma^4$. Since there are many of those, it is convenient to present the result in a form where the trace over flavor indices is already calculated. Then we have
\begin{widetext}
\begin{eqnarray}
\label{G4}
&&\mbox{tr}_{f} \Omega^{(0)} = 
\sum_{i<j} \left\{
\left[\frac{2}{3} (J_{ii}+J_{jj})-\frac{4}{3} J_{ij}\right]
\left(\Gamma_{ij}\Gamma_{ji}\right)^2
+\frac{1}{6} (2J_{ij} -J_{ii}-J_{jj})
\left(\Gamma_{ij}\Gamma_{ij}\right)\left(\Gamma_{ji}\Gamma_{ji}\right) \right\} 
\nonumber \\
&&+\sum_{i\neq j\neq k, \atop j<k} \!\! \left\{
\left[\frac{1}{3}\, (J_{jk}+2J_{iijk})+J_{ii}-J_{ij}-J_{ik}\right]
\left(\Gamma_{ij}\Gamma_{ji}\right)\left(\Gamma_{ik}\Gamma_{ki}\right)
+\left[\frac{1}{3}\, (J_{ii}+2J_{iijk})+J_{jk}-2J_{ijk}\right] 
\left(\Gamma_{ji}\Gamma_{ik}\right)\left(\Gamma_{ki}\Gamma_{ij}\right) 
\right. \nonumber \\
&&+\frac{1}{3}(2J_{iijk}-J_{ij}-J_{ik}) 
\left[ \left(\Gamma_{ii}\Gamma_{jk}\right)\left(\Gamma_{ij}\Gamma_{ki}\right) + 
        \left(\Gamma_{ik}\Gamma_{ji}\right)\left(\Gamma_{kj}\Gamma_{ii}\right)
\right] 
+\left. \frac{1}{3} (2J_{iijk}-J_{ii}-J_{jk})
\left(\Gamma_{ij}\Gamma_{ik}\right)\left(\Gamma_{ki}\Gamma_{ji}\right) 
\right\} \nonumber \\
&&+ \!\! \sum_{i\neq j\neq k}\!\!
\left\{ 
\frac{1}{3} (J_{ik}+2J_{iijk}-J_{ijk}) 
\left[\left(\Gamma_{ii}\Gamma_{ij}\right)\left(\Gamma_{jk}\Gamma_{ki}\right) + 
\left(\Gamma_{ik}\Gamma_{kj}\right)\left(\Gamma_{ji}\Gamma_{ii}\right)\right] 
+ R_{ij} \left[ \left(L^{\alpha\alpha}_- \right)_{ij} \left(\Gamma_{jk}\Gamma_{ki}\right)
                  +\left(\Gamma_{ij}\Gamma_{ji}\right)\left(\Gamma_{ik}\Gamma_{ki}\right)
             \right] 
\right\} \nonumber \\
&&+ \frac{1}{3} \delta_{\alpha\beta\gamma\sigma} 
\sum_{i\neq j}  T_{ij}
\left(\Gamma^\alpha_{ii} \Gamma^\beta_{ii} \Gamma^\gamma_{ij} \Gamma^\sigma_{ji} - 
\Gamma^\alpha_{ii} \Gamma^\beta_{ij} \Gamma^\gamma_{jj} \Gamma^\sigma_{ji}
-\frac{1}{2} \Gamma^\alpha_{ij} \Gamma^\beta_{ji} \Gamma^\gamma_{ij} \Gamma^\sigma_{ji}\right)   \,.
\end{eqnarray}
\end{widetext}
Here, the expression in the parentheses $\left(\Gamma_{ij}\Gamma_{ji}\right)$ is understood to be summed over alpha $\Gamma^\alpha_{ij}\Gamma^\alpha_{ji}$. In the limit of equal quark masses (\ref{G4}) is zero, since all $J$-dependent factors vanish.

\section{Summary}
Let's summarize. Above, we proposed a new efficient method for obtaining the effective action in the theory with heavy particles of unequal masses. Our calculations are based on Eq. (\ref{alg}) which allows us to extend the Schwinger proper-time method to the case of elliptic operators that do not commute with its mass matrix. As a result, we arrive at the effective action (\ref{WE}), in which each of the coefficients can be calculated independently of the other, and corresponds to a certain order of $1/M^2$ expansion. We have explicitly calculated two leading contributions for the case of broken $U(3)\times U(3)$ symmetry. Although we have limited ourselves to this particular example, the formalism is applicable to an arbitrary flavor symmetry group. The method is appropriate for both renormalizable and nonrenormalizable quantum field theories, and has a wide range of applications in various branches of quantum field theory. This includes astrophysics, cosmology, critical phenomena, standard model and beyond, composite Higgs models, etc. 

\section{Acknowledgments}
This work was supported by Grant from Funda\c{c}\~ ao para a Ci\^ encia e Tecnologia (FCT) through the Grant No. CERN/FIS-COM/0035/2019, and the European Cooperation in Science and Technology organisation through the COST Action CA16201 program.

\end{document}